GdCH Nachrichten aus der Chemie

Trendbericht Theoretische Chemie 2020

# Quantendynamik Offener Systeme

Benjamin P. Fingerhut

*Max-Born-Institut für Nichtlineare Optik und Kurzzeitspektroskopie, Berlin, 12489, Germany*

Autoreninformation: Benjamin P. Fingerhut, Max-Born-Institut für Nichtlineare Optik und Kurzzeitspektroskopie, Max-Born-Strasse 2a, 12489 Berlin, Germany.

Phone: +49 30 6392 1404

Email: fingerhut@mbi-berlin.de

**Vorspann**

Quantenmechanik beschreibt die unitäre Zeitentwicklung isolierter Systeme. In der Realität interagiert jedes Quantensystem mit seiner Umwelt, was zu einem irreversiblen Verlust von Phasenbeziehungen führt. Pfadintegralbasierte Methoden erlauben eine rigorose Beschreibung dieser Dekohärenzprozesse.

**DREI FRAGEN AN DEN AUTOR:**

**Welcher Trend ist im letzten Jahr aufgekommen, den Sie so nicht erwartet haben?**

Der technologische Fortschritt der Realisierung von Quantencomputern basierend auf supraleitenden, elektronischen Schaltkreisen hat mich überrascht.

**In welchem Gebiet erwarten Sie in den nächsten 12 Monaten die größten Entwicklungen und warum?**

Durch die Kombination maschinenunterstützter Lerntechniken und neu aufkommender Beschreibungen von Vielkörperquantendynamik erwarte ich Fortschritte in der Simulation dissipativer Quantendynamik in starker Wechselwirkung mit ihrer Umgebung.

**Ihre Forschung in 140 Zeichen?**

Wir entwickeln Methoden zur Simulation nichtadiabatischer Relaxationsdynamik von Schwingungszuständen und elektronisch angeregten Zuständen.

**Konzept und Beispiele Offener Quantensysteme**

Die Eigenschaften quantenmechanischer Systeme werden oft entscheidend durch die Wechselwirkung mit ihrer Umgebung beeinflusst. Die besondere Schwierigkeit liegt hierbei darin, dass das Quantensystem mit einem Vielkörperreservoir wechselwirkt, das einen exponentiell großen Hilbert-Raum besitzt. Da aufgrund dieser Komplexität eine vollständige mikroskopische Beschreibung der Umgebung sich als nahezu unmöglich erweist, ist oftmals ein alternativer Ansatz zielführend. Dabei wechselwirkt ein niederdimensionales System, das quantenmechanisch exakt behandelt wird, als offenes System mit seiner Umgebung, wobei die Umgebung nicht explizit berücksichtigt wird.

Unabhängig von der genauen Ausgestaltung der Umgebung erweist sich die Betrachtung niederdimensionaler Quantensysteme in Kontakt mit ihrer Umgebung als paradigmatisch für die Beschreibung einer Vielzahl wichtigster physikalischer und chemischer Prozesse (siehe Abbildung 1). So werden Protonentransferreaktionen in flüssiger Phase[1] typischerweise über eine Reaktionskoordinate in Form eines Doppelminimumpotentials beschrieben, das stark an die Lösungsmittelfreiheitsgrade gekoppelt ist. Für molekulare Elektronentransferreaktionen und Energietransferprozesse[2,3], die das Herzstück der Photosynthese in biologischen Systemen darstellen, wird die Dynamik durch die Schwingungsfreiheitsgrade des Lösungsmittels bzw. der Enzymumgebung beeinflusst. Quantenmechanische Tunnelprozesse in makroskopischen Zweizustandssystemen[4] stellen die Grundlage für supraleitende QBit-Grundeinheiten aufkommender Quantencomputer dar, in welchen die unvermeidliche Ankopplung an die elektromagnetische Umgebung zu Dekohärenz, also einem Verlust der Phasenbeziehung, führt. In all diesen Szenarien bieten störungstheoretische Methoden einen etablierten Startpunkt, z. B. in Form der weitverbreiteten Markus-, Redfield- oder Förster-Theorie. Diese bieten oftmals aber nur begrenzte Anwendbarkeit[5,6].

Besondere Herausforderungen ergeben sich, wenn das niederdimensionale Quantensystem mit einer strukturierten Umgebung interagiert, wie sie z. B. durch Kernschwingungen induziert wird. Solch strukturierte Umgebungen können ein Gedächtnis für die Wechselwirkung zwischen System und Umgebung verursachen, dessen Zeitskala vergleichbar ist zu Änderungen in der Systemzustandsdynamik. Die Dynamik in diesem Szenario kann besonders komplex sein und solche nicht-Markovschen Gedächtniseffekte können nur schwer durch störungstheoretische Methoden erfasst werden. Die genaue Vorhersage der Zeitentwicklung von nicht-Gleichgewichtszuständen dissipativer Quantensysteme stellt sich somit als ein grundlegendes physikochemisches Problem dar, das insbesondere für strukturierte Umgebungen nach wie vor eine numerisch extrem schwierige Aufgabe ist. So entzieht sich selbst eines der einfachsten Offenen Quantensysteme, ein Zweiniveau-System gekoppelt an ein Wärmebad aus harmonischen Oszillatoren, das sogenannte Spin-Bosonen-Modell, in bestimmten Parameterbereichen einer analytischen oder numerisch exakten Lösung[7].

**Numerische Behandlung**

Die numerisch exakte Simulation Offener Quantensysteme mit System-Bad-Gedächtniseffekten erfolgt im Allgemeinen durch eine Erweiterung des Zustandsraumes, um die Historie des Systems aufgrund der Ankopplung an die Umgebung zu erfassen. Ein leistungsfähiger Ansatz basiert auf der von Feynman vorgeschlagenen Pfadintegralformulierung quantenmechanischer Übergangsamplituden[8]. Komplementäre Methoden stehen mit der *Hierarchical Equation of Motion* (HEOM) genannten Methode[9] und Methoden basierend auf Tensor-Netzwerk-Zuständen (*Tensor Network State* - TNS)[10,11] zur Verfügung. Während die HEOM-Methode auf einer Hierarchie von Bewegungsgleichungen für gekoppelte Dichtematrizen beruht, wird in TNS-Methoden ein Teil der Umgebung innerhalb des System-Hilbert-Raums erfasst, um die relevanten Eigenschaften der vibronischen Wellenfunktion im gesamten System-Bad-Raum darzustellen. Beide Methodenklassen wurden systematisch zur Behandlung komplexer Systeme weiterentwickelt[12-16] (für Übersichtsartikel siehe Ref.[17,18]).

Aus numerischer Sicht stellen sich Pfadintegralmethoden als attraktiv für die Behandlung von Vielkörper-Quantenmechanik dar, da im Gegensatz zu wellenfunktionsbasierten Methoden der mit der Systemgröße exponentiell ansteigende Speicherbedarf umgangen werden kann. Da die Feynman-Pfade eindimensionale Objekte sind, ist deren Zahl unabhängig von der Anzahl der berücksichtigten Freiheitsgrade. Die numerische Summation über alle möglichen Pfade ist jedoch alles andere als trivial. Die Formulierung effizienter Pfadintegrationsmethoden zur Beschreibung Offener Quantensystem nutzt die Tatsache, dass die Freiheitsgrade der Umgebung für eine lineare Ankopplung an ein harmonisches Bad durch Gaussche Integration eliminiert werden können. Die nicht-Markovsche Dynamik wird dann vollständig durch das Einflussfunktional beschrieben, dass die aktuelle Zeitentwicklung an die Historie koppelt und nicht-adiabatische Korrekturen zur Systemdynamik berücksichtigt[19]. Die Methode des quasi-adiabatischen Propagator-Pfadintegrals (*quasi-adiabatic propagator path integral* - QUAPI) nutzt hierbei eine endliche Zeitspanne des nicht-Markovschen Gedächtnisses, um einen generalisierten Dichtematrixtensor iterativ zu konstruieren[20]. Die Annahme einer endlichen Gedächtniszeit ist physikalisch durch die breiten Umgebungsspektren von kondensierter Phase motiviert und stellt formal die einzige Einschränkung der Methode dar. Da die Größe des generalisierten Dichtematrixtensors im Allgemeinen exponentiell mit dem System-Bad-Gedächtniszeitraum skaliert, führt dessen Konstruktion jedoch schnell zu Speicherproblemen. Dies beschränkt die Anwendbarkeit auf Fälle, in denen die System-Bad-Korrelation schnell abklingt (typischerweise kann ein Gedächtniszeitraum von bis zu 20 Propagationszeitschritten berücksichtigt werden) und die spektrale Dichtefunktion keine diskreten Kernfreiheitsgrade beinhaltet.

In jüngster Zeit wurden verschiedene Ansätze verfolgt[7,21,22], um die Behandlung von Langzeit-Gedächtniseffekten in QUAPI-Methoden zu ermöglichen. Die Methode der *time-evolving matrix product operators* (TEMPO)[7,23] verwendet eine Matrix-Produkt-Operator-Darstellung des generalisierten Dichtematrixtensors, die aus der Beschreibung korrelierter niederdimensionaler Systeme adaptiert ist. Durch Singulärwertzerlegung kann der generalisierte Dichtematrixtensor komprimiert und eine polynomiale Skalierung des Rechenaufwands in Simulationen mit langen System-Bad-Korrelationszeiten erreicht werden. Ein alternativer Ansatz basiert auf einem rekursiven Entflechten des generalisierten Dichtematrixtensors[22], das es ermöglicht den exponentiell anwachsenden Speicherbedarf des generalisierten Dichtematrixtensors zu umgehen.

Wir haben kürzlich eine methodische QUAPI-Variante entwickelt, die auf einer intermediären, groben Darstellung des Einflussfunktionals beruht (*mask assisted coarse graining of influence coefficients* - MACGIC-QUAPI)[21]. Die Methode nützt die physikalischen Eigenschaften der Badkorrelationsfunktion, die zu frühen Zeiten durch ein schnelles Abklingen und zu späten Zeiten durch eine langsame zeitliche Variation charakterisiert ist. Dieses Verhalten ermöglicht es das Kurzzeit-System-Bad-Gedächtnis genau (also durch eine größere Anzahl von Gitterpunkten), das Langzeit-System-Bad-Gedächtnis jedoch durch eine grobe, nicht-äquidistante Darstellung zu charakterisieren. Somit lässt sich die Gesamtzeitdauer des System-Bad-Gedächtnisses von der Anzahl der berücksichtigten Feynman-Pfade entkoppeln. Das Verfahren eröffnet den numerischen Zugang zu langen Badkorrelationszeiten und ermöglicht die exakte Simulation komplexerer Systeme und Umgebungen.

Exemplarisch ist in Abbildung 2 die Zeitentwicklung dissipativer Quantendynamik dargestellt, wie sie in einem Modell des bakteriellen Reaktionszentrums abläuft[24]. Die MACGIC-QUAPI-Methode ermöglicht es ausgehend von exzitonisch kohärenter Dynamik die irreversible Fixierung von Energie in einem ladungsseparierten Zustand zu untersuchen. Exzitonengekoppelte Ladungstransferdynamik in Reaktionszentren stellt einen der grundlegenden Mechanismen für die Nutzung von Sonnenenergie in biologischen Systemen und synthetisierten organischen Materialien dar und ist oftmals durch eine starke Wechselwirkung zwischen elektronischen und Kernfreiheitsgraden charakterisiert. Dies erfordert nicht-störungstheoretische Simulationen unter Verwendung strukturierter spektraler

Dichtefunktionen. Die rigorose theoretische Beschreibung stellt also nach wie vor eine besondere Herausforderung dar. Die dargestellte Modelierung der im Reaktionszentrum ablaufenden Ladungstransferdynamik deckt komplexe Dynamik ab, die von anfänglich kohärenter Oszillationsdynamik bis hin zum inkohärenten „Einfangen" der Population im ladungsgetrennten Zustand reicht. Letzterer wird durch eine starke Ankopplung an die Umgebung charakterisiert. Die hohe numerische Effizienz der MACGIC-QUAPI-Methode erlaubt es außerdem den Einfluss von Schwingungsmoden auf die kohärente Populationsdynamik und deren Persistenz im Modell des bakteriellen Reaktionszentrums zu untersuchen. Die Ergebnisse deuten auf einen moderaten Einfluss der Schwingungsmoden auf die Ladungstransferdynamik hin, während die kohärent ablaufende exzitonische Dynamik durch Resonanzen der Kernfreiheitsgrade stark beeinflusst wird. Es liegt also eine Entkopplung der Zeitskalen des kohärenten Anregungsenergietransfers und der Ladungstransferdynamik vor. Unsere Ergebnisse unterstützen hierbei ein Bild, in dem molekulare Schwingungen optimalen, nicht aktivierten Ladungstransfer durch die Gewährleistung von Resonanzbedingungen sicherstellen und diesen gegen Störungen abschirmen.

**Zusammenfassung und Ausblick**

Methodische Fortschritte in der Beschreibung der Echtzeitdynamik Offener Quantensysteme erlauben Untersuchungen immer komplexerer Systeme und Umgebungen. Dies eröffnet, basierend auf akkuraten ab initio Parametrisierungen der System- und Badfreiheitsgrade[25-27], die Möglichkeit mikroskopischer Modellierung von Dynamik wie sie in makromolekularen biologisch und technologisch relevanten Systemen abläuft. Während der Einfluss der Umgebungsfluktuationen auf die Systemenergetik (diagonale System-Bad-Wechselwirkung) zunehmend verstanden ist und in dynamischen Simulationen zuverlässig berücksichtigt werden kann, ist der Einfluss der Umgebung auf den Transferprozess durch nicht-diagonale Ankopplung an das Quantensystem weniger untersucht. Die Behandlung solcher, nicht diagonaler System-Bad-Wechselwirkung[28] und die Berücksichtigung mehrerer, nicht-kommutierender Bäder[29] stellen nach wie vor Herausforderungen für die numerische Beschreibung Offener Quantensysteme dar. Einblick in die durch nicht-Markovsche Gedächtniseffekte komplex ablaufende Dynamik kann neue Möglichkeiten der Kontrolle und Optimierung technologisch wichtiger Energie- und Ladungstransferprozesse ermöglichen, die sich erst aus der Vielkörperphysik komplexer Systeme ergeben. Derart tiefes Verständnis ist nur durch numerische Simulationsmethoden zu erlangen, die erheblich über einfache Wärmebadbeschreibungen der Umgebung hinausgehen.

Benjamin P. Fingerhut

Max-Born-Institut für Nichtlineare Optik und Kurzzeitspektroskopie

fingerhut@mbi-berlin.de8

**Literatur:**

**Abbildungslegenden:**

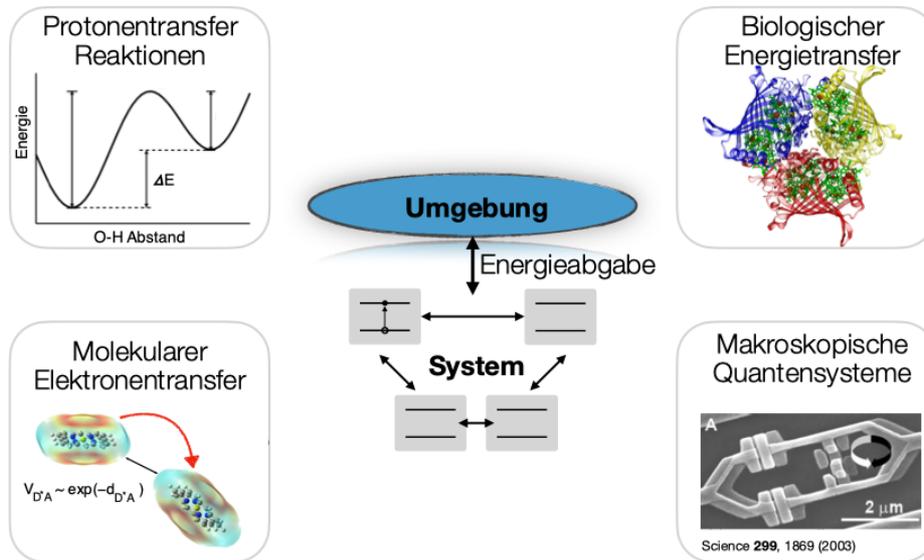

**Abbildung 1**. Exemplarische Beispiele niederdimensionaler Quantensysteme, deren Dynamik maßgeblich durch die Wechselwirkung mit der Umgebung geprägt wird.

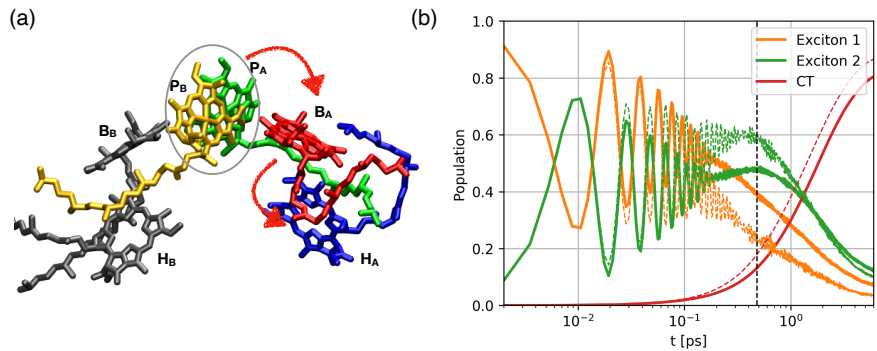

**Abbildung 2**. Dissipative Quantendynamik in einem Modell des bakteriellen Reaktionszentrums (24): nicht-Gleichgewichtsdynamik wird am *Special Pair* ($P_B P_A$) initiiert (a) und weist aufgrund starker exzitonischer Kopplung kohärente Zeitentwicklung auf (b). Die Wechselwirkung mit der Umgebung und molekulare Elektronentransferprozesse ($P_B P_A \rightarrow B_A \rightarrow H_A$) verursachen Dephasierungsprozesse und führen zur Ausbildung des ladungsseparierten Zustands. Irreversibilität der Ladungstrennseparation erfolgt auf der Nanometer-Größen- und Pikosekunden-Zeitskala.

**Kurzbiographie**

Dr. Benjamin Fingerhut promovierte an der Ludwig-Maximilians-Universität München. Er trat 2014 dem Max-Born-Institut bei und leitet dort die Nachwuchsgruppe Biomolekulare Dynamik. Die Gruppe untersucht ultraschnelle strukturelle Dynamik molekularer und biomolekularer Systeme. Seine Forschung wird seid 2019 durch einen ERC Starting Grant des Europäischen Forschungsrats unterstützt.